\documentclass[aps,twocolumn,prb,preprintnumbers,amsmath,amssymb]{revtex4-1}
\allowdisplaybreaks[4]
\usepackage{here}
\newcommand{\lsim} 
 {\ \raise.35ex\hbox{$<$}\kern-0.75em\lower.5ex\hbox{$\sim$}\ }
\newcommand{\gsim}
 {\ \raise.35ex\hbox{$>$}\kern-0.75em\lower.5ex\hbox{$\sim$}\ }

\usepackage{graphicx}
\usepackage{dcolumn}
\usepackage{bm}
\usepackage{amsmath}
\usepackage{times}
\usepackage[usenames]{color}
\usepackage{natbib}
\usepackage{ulem}

\begin{document}
\title{Anomalous Hall effect in $\kappa$-type organic antiferromagnets}
\author{Makoto Naka$^{1}$, Satoru Hayami$^{2}$, Hiroaki Kusunose$^{3}$, Yuki Yanagi$^{4}$, Yukitoshi Motome$^2$, and Hitoshi Seo$^{5,6}$}
\affiliation{$^1$Waseda Institute for Advanced Study, Waseda University, Shinjuku, Tokyo 169-8050, Japan}
\affiliation{$^2$Department of Applied Physics, The University of Tokyo, Bunkyo, Tokyo 113-8656, Japan}
\affiliation{$^3$Department of Physics, Meiji University, Kawasaki, Kanagawa 214-8571, Japan}
\affiliation{$^4$Institute for Materials Research, Tohoku University, Sendai, Miyagi 980-8577, Japan}
\affiliation{$^5$Condensed Matter Theory Laboratory, RIKEN, Wako, Saitama 351-0198, Japan}
\affiliation{$^6$Center for Emergent Matter Science (CEMS), RIKEN, Wako, Saitama 351-0198, Japan}
\date{\today}
\begin{abstract} 
We theoretically propose a mechanism for the anomalous Hall effect (AHE) in an antiferrromagnetic (AFM) state of $\kappa$-type organic conductors. 
We incorporate the spin-orbit coupling in the effective Hubbard model on the $\kappa$-type lattice structure taking into account the orientation of the molecules and their arrangement with dimerization. 
Treating this model by means of the Hartree-Fock approximation and the linear response theory, we find that an intrinsic contribution to the Hall conductivity becomes nonzero in the electron-doped AFM metallic phase with a small canted ferromagnetic moment.
We show that, contrary to the conventional wisdom, the spin canting is irrelevant to the Hall response; the nonzero Hall conductivity originates from the collinear component of the AFM order in the presence of the spin-orbit coupling.
These features are well explained analytically in the limit of strong dimerization on the anisotropic triangular lattice. 
Furthermore, we present an intuitive picture for the present AHE by considering the real-space configuration of emergent magnetic fluxes.
We also find that the Hall response appears even in the undoped AFM insulating phase at nonzero frequency as the magneto-optical Kerr effect, which is enhanced around the charge transfer excitations.
We discuss possible detections of the AHE in ET based compounds.
\end{abstract} 


\maketitle
\narrowtext



%
%

%


\section{Introduction}
The anomalous Hall effect (AHE) is one of the long standing issues in condensed matter physics, studied for over a century.~\cite{hall, nagaosa} 
The AHE originates from an effective internal magnetic field emerging from the spin-orbit coupling (SOC) and magnetic spin structures breaking time-reversal symmetry.~\cite{karplus} 
In light of the guiding principle, not only ferromagnets but also antiferromagnets with noncollinear spin textures giving rise to an effective (or fictitious) magnetic field have been extensively studied for the past decades.~\cite{ohgushi, shindou, ohgushi2, tomizawa, chen, nakatsuji} 
More recently, even antiferromagnets with collinear spin structures are proposed to show an AHE.~\cite{smejkal, li, feng} 
In such systems, the crystal lattice symmetry under the antiferromagnetic (AFM) spin ordering is essential for the AHE, while its microscopic mechanism is not fully elucidated. 
For example, it is unclear how we can achieve an intuitive picture of the electrons feeling the Lorentz force by the effective magnetic field as discussed in the ferromagnetic (FM) and noncollinear AFM cases.

Recently, the authors revealed that, in an organic antiferromagnet $\kappa$-(ET)$_2X$,~\cite{note4} a collinear AFM order yields a spin dependent band splitting and a spin current generation even in the absence of the SOC.~\cite{naka} 
This unconventional phenomenon comes from the breaking of glide symmetry in the molecular arrangement by the AFM order.~\cite{hayami1, hayami2, hayami3}
In fact, when taking into account the SOC, such a molecular degree of freedom can potentially provide another platform for the AHE. 
Here we present a theory of the AHE in $\kappa$-(ET)$_2X$, which requires neither a net FM moment nor a noncollinear magnetic spin structure. 
We will show that, in contrast to the conventional mechanisms, the AFM ordering and the SOC, both under influence of the underlying molecular arrangement, are the key ingredients for the appearance of the AHE. 
\begin{figure*}[t]
\begin{center}
\includegraphics[width=1.5\columnwidth, clip]{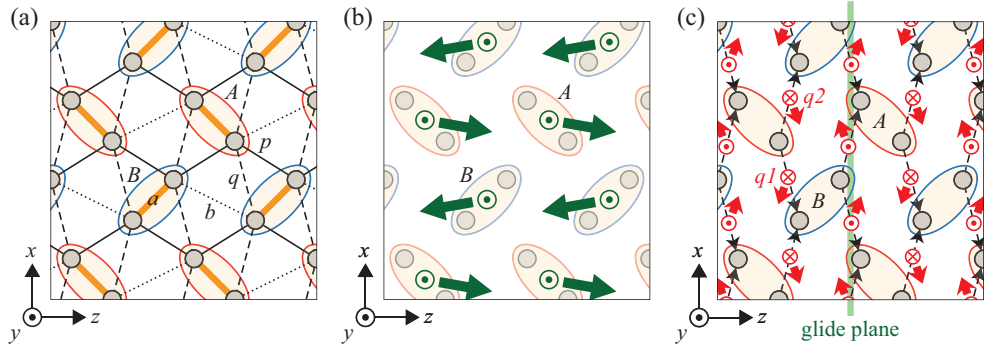}
\end{center}
\caption{
(a) Schematic two-dimensional molecular arrangement in $\kappa$-(ET)$_2X$.
The circles and ellipses represent the ET molecules and dimers, respectively. 
$A$ and $B$ stand for the two dimers in the unit cell with different orientations.
The thick solid, thin solid, broken, and dotted lines, denoted by $a$, $p$, $q$, and $b$, respectively, are the intermolecular bonds, on which the transfer integrals in the model [Eq.~(\ref{hHubb})] are defined.
(b) Schematic illustration of the canted AFM spin structure. 
The green arrows and circled dots show the directions of the local spin moments on the dimers by their components parallel and perpendicular to the $zx$ plane, respectively. 
(c) Spatial distribution of the antisymmetric SOC vector ${\bm \lambda}_{ij}$.
The red solid arrows together with their $y$ components indicate the directions of ${\bm \lambda}_{ij}$ on the $q$ bonds associated with the electron hopping along the directions of the black broken arrows.
The green line denotes the glide plane perpendicular to the $zx$ plane. 
${\bm \lambda}_{ij}$ on the $q$ bonds, $q1$ and $q2$, are connected by the glide symmetry.
}
\label{fig1}
\end{figure*}

The crystal structure of $\kappa$-(ET)$_2X$ consists of an anisotropic triangular lattice of dimers of ET molecules with two kinds of orientations, termed $A$ and $B$, as shown in Fig.~\ref{fig1}(a).~\cite{miyagawa}
We call this the $\kappa$-type molecular arrangement in the following.
The ET layers are stacked alternately with the insulating anion $X$ layers.
The frontier molecular orbitals in each ET dimer become hybridized by the intradimer transfer integral and constitute bonding and antibonding orbitals. 
In these two orbitals, there are three electrons per dimer on average, and then the energy bands originating from them are three-quarter filled. 
When the dimerization is large one can regard the system as effectively half-filled of the antibonding orbital bands.~\cite{kino} 
Therefore, owing to the electron-electron interaction, $\kappa$-(ET)$_2X$ locates on the verge of the Mott metal-insulator transition.~\cite{miyagawa, lefebvre, limelette, kagawa2}

In the Mott insulating phase, an AFM spin ordering takes place in most of the $\kappa$-type compounds at low temperatures, where the spins on the $A$ and $B$ dimers form an almost collinear AFM order with small canting.~\cite{miyagawa2} 
The canting originates from the fact that there is no inversion center on the bonds connecting the $A$ and $B$ dimers because of their molecular orientations [see Fig.~\ref{fig1}(a)]; the SOC generates the Dzyaloshinskii-Moriya (DM) interaction that twists the spins.~\cite{moriya}

The canted AFM spin structure and the arrangement of the DM vectors in this system have been controversial for a long time because of the difficulty of neutron diffraction experiments in organic compounds.~\cite{welp, pinteric, smith, kagawa}
Recently, Ishikawa {\it et al.} have determined the AFM structure in $\kappa$-(ET)$_2$Cu[N(CN)$_2$]Cl by combining detailed magnetization measurements and calculations for the classical Heisenberg model with the empirical DM interaction.~\cite{ishikawa} 
The proposed AFM structure in the ET layer is schematically shown in Fig.~\ref{fig1}(b). 
The AFM moment is almost parallel to the $z$ axis and the weak FM moment is in the $xy$ plane.
Here we take the coordinate axes referring to the crystal axes; the $x$ and $z$ axes are set along the interlayer $a$ and $c$ axes, and the stacking direction is the $y$ direction.~\cite{note3}
Also, the SOC in this system has recently been investigated from the theoretical side.
Winter {\it et al.} estimated the SOC and DM vectors in a series of $\kappa$-type ET compounds by means of {\it ab initio} quantum chemical calculations,~\cite{winter} whose results agree with those by Ishikawa {\it et al}.

In this paper, we theoretically study the AHE under the combination of the AFM ordering and the SOC in $\kappa$-(ET)$_2X$. 
Considering a Hubbard-type tight-binding model with the SOC, we obtain the ground-state phase diagram by the Hartree-Fock (HF) approximation, and calculate the Hall conductivity and optical responses by the linear response theory.
At three-quarter filling, the canted AFM insulating phase with a small FM moment is realized as the ground state, reproducing the recent experiment.
Although the DC Hall conductivity is zero in the AFM insulating phase, we find that it becomes nonzero, i.e., the AHE appears, when the electrons are doped leading to the AFM metallic phase. 
In order to pin down the mechanism of the AHE, we construct an effective model in the strong dimerization limit, and derive the analytic expression of the Hall conductivity. 
The formula clearly shows that the AHE relies on not the small FM moment associated with the spin canting but a cooperative effect of the collinear AFM ordering and a fictitious magnetic field emerging from the SOC.
On the other hand, in the AFM insulating phase the AHE appears in the transverse optical response, i.e., as the magneto-optical Kerr effect, which shows nonzero oscillator strength and rotation angle for the mid-infrared light in the frequency range between the interdimer and intradimer charge transfer excitations. 
Finally, we propose how to verify our proposal in the $\kappa$-type ET compounds.

\section{Model and Method}
The Hubbard model based on the frontier molecular orbitals in the $\kappa$-type ET system is given by~\cite{kino} 
\begin{align}
{\cal H}_{\rm Hubb} = \sum_{ij} \sum_{\sigma} t_{ij} c^{\dagger}_{i \sigma} c_{j \sigma} + U \sum_{i} n_{i \uparrow} n_{i \downarrow},
\label{hHubb}
\end{align}
where $c_{i \sigma}$ ($c_{i \sigma}^\dagger$) and $n_{i \sigma}(=c^\dagger_{i \sigma}c_{i \sigma})$ are the annihilation (creation) and number operators of an electron at $i$th ET molecule with spin $\sigma$, respectively. 
$t_{ij}$ represents the intermolecular transfer integrals $(t_a, t_p, t_q, t_b)$ on the bonds shown in Fig.~\ref{fig1}(a) and $U$ is the intramolecular Coulomb interaction. 
The SOC Hamiltonian of this system is described by complex electron transfer integrals depending on the spins as
\begin{align}
{\cal H}_{\rm SOC} = \sum_{ij} \sum_{\sigma \sigma'} \frac{i}{2} ({\bm \lambda}_{ij} \cdot {\bm \sigma})_{\sigma \sigma'} c^{\dagger}_{i \sigma} c_{j \sigma'}, 
\end{align}
where $\bm \sigma$ is the vector of Pauli matrices and the vector ${\bm \lambda}_{ij}(=-{\bm \lambda}_{ji})$ is the antisymmetric SOC vector arising from the second-order perturbation in terms of the multiorbital intermolecular hoppings and the atomic SOC of ET molecules.~\cite{moriya, winter}
Among the four kinds of intermolecular bonds in Fig.~\ref{fig1}(a), ${\bm \lambda}_{ij}$ is nonzero only on the $p$ and $q$ bonds on which the local inversion center is absent.
The spatial distribution of ${\bm \lambda}_{ij}$ on the $q$ bonds is illustrated in Fig.~\ref{fig1}(c), which are much larger than those on the $p$ bonds according to Ref.~[\onlinecite{winter}]. 
There are two ${\bm \lambda}_{ij}$ on the $q$ bonds, which we denote as ${\bm \lambda}_{q1}$ and ${\bm \lambda}_{q2}$. 
They are connected by the glide symmetry with respect to the $zx$ plane, namely, $({\lambda}_{q2}^x, \lambda_{q2}^y, \lambda_{q2}^z) = (\lambda_{q1}^x, \lambda_{q1}^y, -\lambda_{q1}^z)$; the same holds for the $p$ bonds.

\begin{figure}[t]
\begin{center}
\includegraphics[width=0.7\columnwidth, clip]{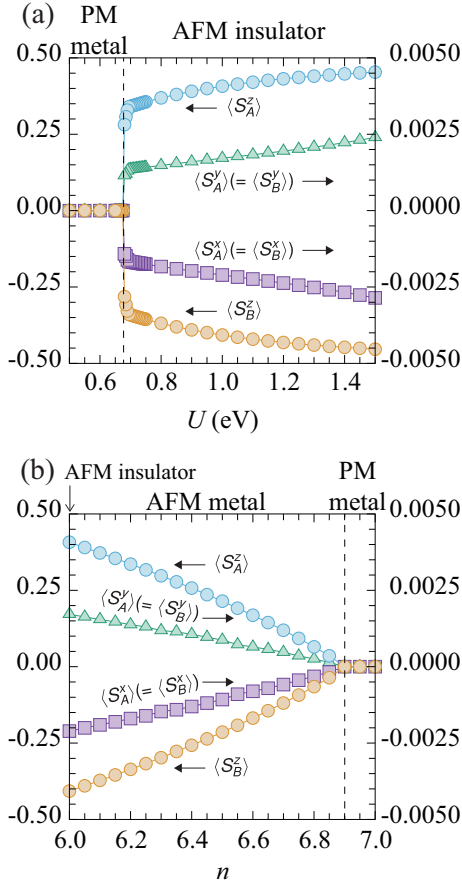}
\end{center}
\caption{
(a) $U$ dependence of each component of the expectation value of the spin moments $\langle {\bm S}_{X} \rangle$ on the $X$ ($=A, B$) dimers at $n=6$.
(b) $n$ dependence of $\langle {\bm S}_{X} \rangle$ at $U=1$ eV.
}
\label{fig2}
\end{figure}
\begin{figure}[t]
\begin{center}
\includegraphics[width=1.0\columnwidth, clip]{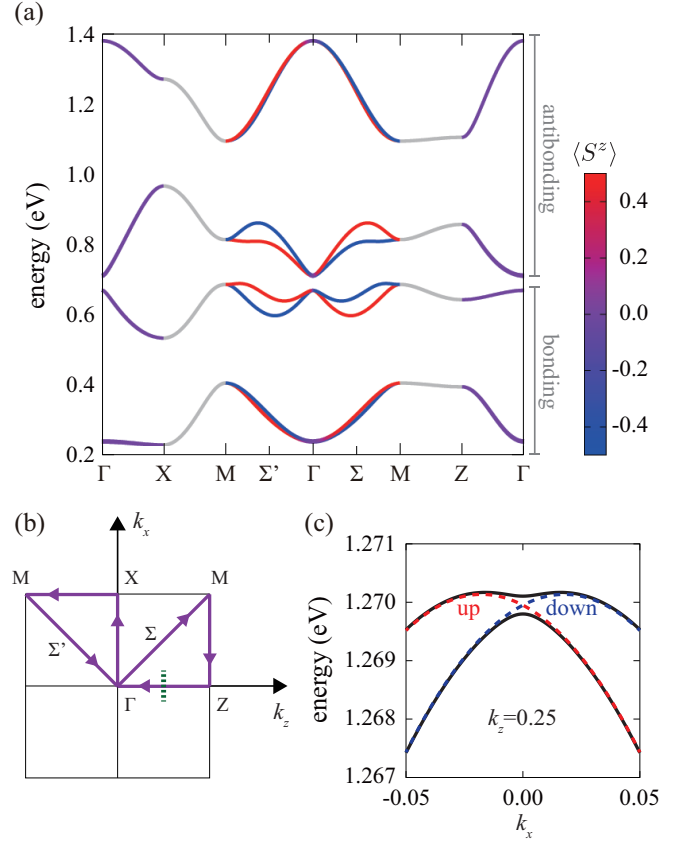}
\end{center}
\caption{
(a) Energy band structure in the AFM insulating state at $(U, n)=(1 \ \rm{eV}, 6)$.
The colors of the bands indicate the magnitude of the expectation value of $S^z$ for their Bloch states.
The gray lines represent the spin degenerate bands on the zone boundaries (X-M and M-Z) where $\langle S^z \rangle$ is not uniquely determined.
(b) The trajectory of the symmetric lines in the first BZ for the energy bands in (a).
(c) The dispersions of the top two bands in (a) along the green broken line in (b).
The solid and broken lines are the energy dispersions with and without the SOC, respectively.
$k_z$ and $k_x$ stand for the coefficients of the reciprocal vectors ${\bm b}_{z}$ and ${\bm b}_{x}$, respectively.
}
\label{fig3}
\end{figure}
We analyze the total Hamiltonian ${\cal H} = {\cal H}_{\rm Hubb} + {\cal H}_{\rm SOC}$ within the HF approximation, where the interaction term is decoupled as $n_{i \uparrow} n_{i \downarrow} \rightarrow n_{i \uparrow} \langle n_{i \downarrow} \rangle + \langle n_{i \uparrow} \rangle n_{i \downarrow} - \langle n_{i \uparrow} \rangle \langle n_{i \downarrow} \rangle - c^\dagger_{i \uparrow} c_{i \downarrow} \langle c^\dagger_{i \downarrow} c_{i \uparrow} \rangle - \langle c^\dagger_{i \uparrow} c_{i \downarrow} \rangle c^\dagger_{i \downarrow} c_{i \uparrow} + \langle c^\dagger_{i \uparrow} c_{i \downarrow} \rangle \langle c^\dagger_{i \downarrow} c_{i \uparrow}\rangle$.
We take the unit cell including the two neighboring dimers $A$ and $B$ and determine the mean fields self-consistently in the ground state.

We calculate the transport properties by the linear response theory.
Within the HF approximation, the total electric current operator is defined by
\begin{align}
{\bm J} = \frac{1}{i \hbar} \left[ {\bm P}, {\cal H}_{\rm HF} \right],  
\end{align}
where ${\cal H}_{\rm HF}$ represents the HF Hamiltonian and ${\bm P}$ is the electric polarization operator defined by ${\bm P} = - e \sum_{i} n_{i} {\bm r}_i$, where ${\bm r}_i$ is the position vector of the $i$th ET molecular site.
Using the Kubo formula, the electric conductivity along the $\mu$ axis with respect to an electric field parallel to the $\nu$ axis ($\mu, \nu = x, z$) is given by 
\begin{align}
\sigma_{\mu \nu}(\omega) = \frac{\hbar}{iN a_z a_x} \sum_{{\bm k}lm} &\frac{f(\epsilon_{{\bm k}l}) - f(\epsilon_{{\bm k}m})}{\epsilon_{{\bm k}l} - \epsilon_{{\bm k}m}} \notag \\
&\times \frac{[J^{\mu}(\bm k)]_{ml} [J^{\nu}(\bm k)]_{lm}}{\hbar \omega + \epsilon_{{\bm k}m} - \epsilon_{{\bm k}l} + i\gamma},
\label{sigma}
\end{align}
where $f(\epsilon_{{\bm k}l})$ is the Fermi distribution function for the Bloch eigenstate of ${\cal H}_{\rm HF}$ with wave vector $\bm k$ and band index $l$.
$[J^{\mu}(\bm k)]_{ml}$ is the matrix element of the $\mu$ component of the total electric current operator between these Bloch eigenstates, $\omega$ is the frequency of the external electric field, and $\gamma$ is the damping constant; $a_z$ and $a_x$ are the lattice constants for the $z$ and $x$ directions, respectively, and $N$ is the total number of unit cells.
We define the real and imaginary parts of the Hall conductivity as $\sigma_{\mu\nu}(\omega) = \sigma'_{\mu\nu}(\omega) + i \sigma''_{\mu\nu}(\omega)$.

\begin{figure*}[t]
\begin{center}
\includegraphics[width=2.0\columnwidth, clip]{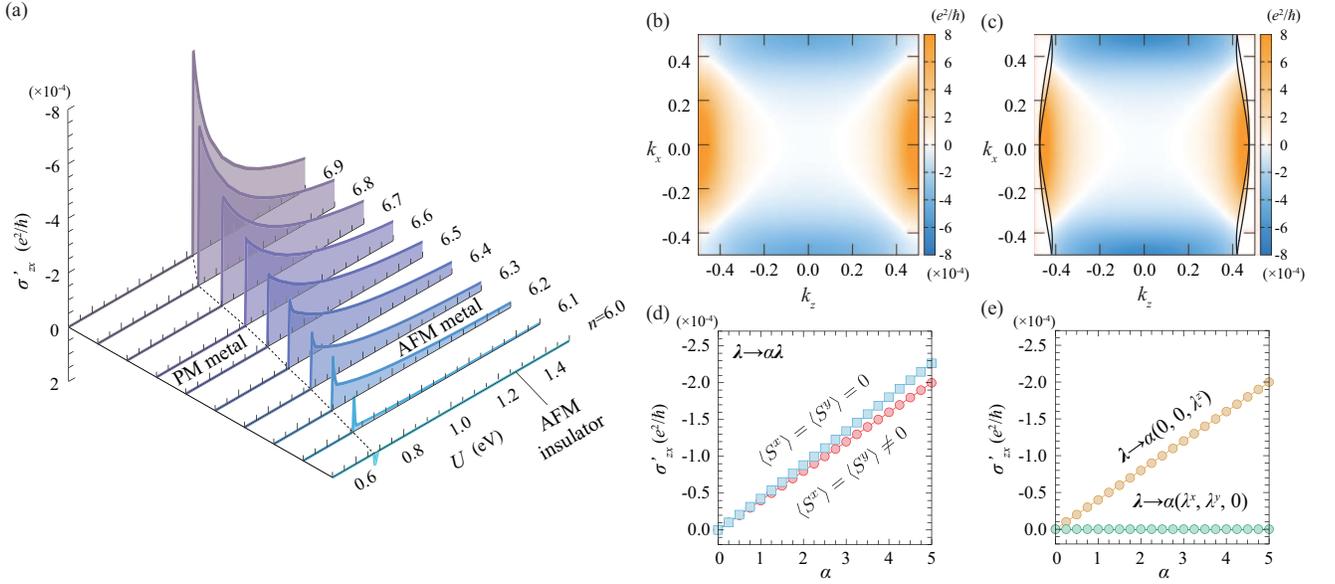}
\end{center}
\caption{
(a) Real part of the DC Hall conductivity, $\sigma_{zx}'$, in the ground state as a function of the intramolecular Coulomb interaction $U$ and the number of electrons per unit cell, $n$. 
The broken line on the basal plane represents the phase boundary between the PM and AFM phases. 
(b) and (c) Color maps of the ${\bm k}$-resolved Hall conductivity in the BZ (b) in the AFM insulating phase at $(U, n) = (1 \ {\rm eV}, 6)$ and (c) in the AFM metallic phase at $(U, n) = (1 \ {\rm eV}, 6.2)$.
$k_z$ and $k_x$ stand for the coefficients of the reciprocal vectors ${\bm b}_{z}$ and ${\bm b}_{x}$, respectively.
The solid lines in (c) represent the Fermi surfaces of the spin-split bands. 
The damping constant is fixed at $\gamma = 0.001$~meV.
(d) Variation of $\sigma_{zx}'$ with respect to artificially tuned SOC multiplying the original ${\bm \lambda}_{ij}$ by $\alpha$.
The circle and square symbols show the results in the canted AFM state and the collinear AFM state, respectively. 
(e) Variation of $\sigma_{zx}'$ when the $z$ or $xy$ components of $\alpha {\bm \lambda}_{ij}$ are set to zero.
The parameters are fixed at $(U, n) = (1 \ {\rm eV}, 6.2)$ in (d) and (e).
}
\label{fig4}
\end{figure*}
We adopt the values of the transfer integrals from a first-principles band calculation~\cite{koretsune} as $(t_a, t_p, t_q, t_b) = (-0.207, -0.102, 0.043, -0.067)$~eV, and the SOC vectors from the quantum chemical calculation~\cite{winter} as ${\bm \lambda}_{p1} = (\lambda_{p1}^x, \lambda_{p1}^y, \lambda_{p1}^z) =(-0.3, 0.12, 0.1)$~meV, ${\bm \lambda}_{p2} = (-0.3, 0.12, -0.1)$~meV, ${\bm \lambda}_{q1} = (-0.88, -0.99, -0.18)$~meV, ${\bm \lambda}_{q2} = (-0.88, -0.99, 0.18)$~meV. 
We take the damping constant as $\gamma = 1$~meV unless otherwise noted. 
We calculate the ground-state properties changing the intramolecular Coulomb interaction $U$ and the number of electrons per unit cell, $n$, as the parameters. 
The $\bm k$-space mesh ($=N$) is chosen as $200 \times 200$ and $1000 \times 1000$ for evaluating the order parameters and the Hall conductivity, respectively.
Since the Hall conductivity is antisymmetric as $\sigma_{zx}(\omega) = - \sigma_{xz}(\omega)$, we present only $\sigma_{zx}(\omega)$ in the following.

\section{Results}
\subsection{AFM order and band structure}
First, we examine the spin structure in the ground state.
When the SOC is absent, previous studies show that a collinear AFM order with opposite spin directions on the $A$ and $B$ dimers is stabilized when $U$ is increased.~\cite{kino, naka} 
Our results show that the SOC induces spin canting from this, whose AFM structure is consistent with the experiments. 
Figure~\ref{fig2}(a) shows the spin moments $\langle {\bm S}_{X} \rangle$ ($X=A, B$) on each dimer in the unit cell at three-quarter filling ($n=6$) as a function of the intramolecular Coulomb interaction $U$.
With increasing $U$, the system undergoes a phase transition from the paramagnetic (PM) metallic phase to the AFM insulating phase at around $U=0.68$ eV. 
In the latter, the AFM moment is almost parallel to the $z$ axis and a small FM moment appears in the $xy$ plane as in Fig.~\ref{fig1}(b); namely, an almost collinear AFM is stabilized. 
Note that the canting is very small due to the small SOC: The transverse spin component is about $10^2$ times smaller than the longitudinal one.
The directions of the AFM easy axis and the FM moment well reproduce the results of the recent magnetization measurement on $\kappa$-(ET)$_2$Cu[N(CN)$_2$]Cl.~\cite{ishikawa}

On the other hand, when $n$ is increased from the three-quarter filling, the AFM insulating state immediately turns metallic.~\cite{note} 
The electron-doped state retains the canted AFM spin structure, whose AFM and FM moments, however, decrease monotonically and vanish simultaneously at a critical value of $n$ that depends on $U$ [see also Fig.~\ref{fig4}(a)]. 
The behavior is exemplified for $U=1$~eV in Fig.~\ref{fig2}(b).
\begin{figure*}[t]
\begin{center}
\includegraphics[width=1.5\columnwidth, clip]{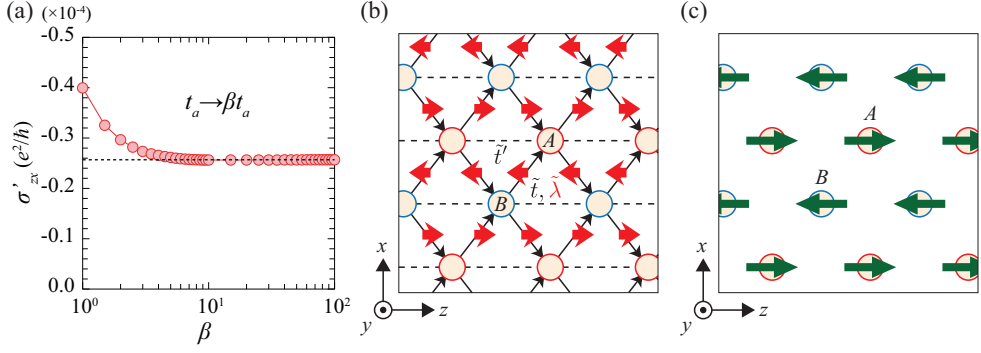}
\end{center}
\caption{
$t_a$ dependence of $\sigma_{zx}'$ at $(U, n) = (1 \ {\rm eV}, 6.2)$. 
The broken line shows the Hall conductivity calculated by the analytic formula [Eq.~(\ref{sigma_tilde})] for the effective model in the strong dimerization limit ($t_a \rightarrow \infty$).
(b) Lattice structure of the effective model. 
The circles represent the $A$ and $B$ dimers that are the lattice sites on the anisotropic triangular lattice.
The solid and broken lines represent the transfer integrals $\tilde{t}$ and $\tilde{t}'$, respectively. 
The red arrows represent the Zeeman-type SOC vectors $\tilde{\bm \lambda}_{ij}$, which is nonzero only in the $z$ component, associated with the electron hoppings denoted by the black arrows. 
(c) The collinear AFM spin structure on the anisotropic triangular lattice. 
The green arrows represent the directions of the spin moments.
}
\label{fig5}
\end{figure*}

Figure~\ref{fig3}(a) shows the energy band structure in the AFM insulating phase at $(U, n)=(1 \ {\rm eV}, 6)$. 
The symmetric lines in the first Brillouin zone (BZ) are indicated in Fig.~\ref{fig3}(b).
As there are four independent ET molecules in the unit cell, the number of the energy bands in the first BZ is $8=4 \times 2$ taking into account of the spin degree of freedom.
The lower and upper four bands correspond to the bonding and antibonding bands, respectively. 
Both of them are further separated into two each by energy gaps due to the AFM ordering. 
The Fermi energy is located in the AFM gap of the antibonding band around $1$~eV.
The degeneracy with respect to the spin degree of freedom is lifted in the whole BZ except for the zone boundaries.
Then, the direction of the spin for each Bloch state is locked at each $\bm k$ point (spin-momentum locking); the $\bm k$-space variation of the $S^z$ component is illustrated as the color of each band in Fig.~\ref{fig3}(a).
The large spin splitting emerging on the $\Gamma$-M lines is caused by the mechanism found in our previous work which is present even without the SOC,~\cite{naka} i.e., the cooperative effect of the AFM ordering and the ET molecular orientations. 
On the other hand, the small spin splitting on the $k_z$ and $k_x$ axes is attributable to the presence of the SOC.
Figure~\ref{fig3}(c) shows the energy dispersions of the top bands in Fig.~\ref{fig3}(a) with and without the SOC along the path denoted by the broken line in the BZ shown in Fig.~\ref{fig3}(b).
The degenerate $S^z$-up and $S^z$-down bands on the $k_z$ axis are hybridized by the SOC, resulting in the spin splitting of the order of $\left| {\bm \lambda}_{ij} \right|$.

\subsection{AHE}
Next, we investigate the Hall conductivity in the ground state. 
Figure~\ref{fig4}(a) shows the real part of the DC Hall conductivity, $\sigma_{zx}'(\omega=0)$, as a function of $U$ and $n$.
The broken line in the basal plane denotes the phase boundary between the PM and AFM phases; the critical value of $U$ for the appearance of the AFM order increases as $n$ deviates from $6$. 
The Hall conductivity is zero in both the AFM insulating phase at $n=6$ and the PM metallic phase for all $n$, while it turns finite in the AFM metallic phase in the electron-doped region. 
One can see that the magnitude of $\sigma_{zx}'$ increases with the decrease (increase) of $U$ ($n$), corresponding to decrease in the AFM order parameter (see Fig.~\ref{fig2}).

In order to understand the reason why the AHE appears in the electron-doped AFM metallic region, we decompose the Hall conductivity in $\bm k$ space. 
Figure~\ref{fig4}(b) shows the $\bm k$-resolved Hall conductivity defined by $\sigma'_{zx} \equiv e^2/(N \hbar)\sum_{\bm k} \sigma'_{zx}({\bm k})$ in the AFM insulating phase.
We note that $\sigma'_{zx}({\bm k})$ is connected to the $y$ component of the Berry curvature of the $m$th band, $b^y_m({\bm k})$, as $\sigma'_{zx}({\bm k}) = - \sum_{m} f(\epsilon_{{\bm k}m}) b^y_{m}({\bm k})$.~\cite{nagaosa} 
$\sigma'_{zx}({\bm k})$ exhibits a $d$-wave-like sign change centered at the $\Gamma$ point, whose $\bm k$ dependence can be approximated as $\sin [\pi (k_z - k_x)] \sin [\pi (k_z + k_x)]$. 
In this case, in the AFM insulating phase where the Fermi surface is absent, $\sigma'_{zx}$ given by the summation over the whole BZ is zero by the cancellation between the positive and negative contributions. 
On the other hand, in the AFM metallic phase, this cancellation becomes incomplete since the contributions from outside of the Fermi surfaces in the $\bm k$ space, where all the Bloch states are fully occupied, vanish due to the Fermi distribution functions in Eq.~(\ref{sigma}); see Fig.~\ref{fig4}(c).
This leads to the nonzero Hall conductivity in the AFM metallic phase.

The present AHE originates from an intrinsic mechanism independent of the damping factor $\gamma$.
In fact, we find that the value of $\sigma'_{zx}$ becomes nearly constant for $\gamma \lesssim 0.1$ eV, which is the order of the AFM gap. 

To figure out the relevant elements for the AHE, we first artificially tune the SOC and investigate how $\sigma'_{zx}$ varies. 
Figure~\ref{fig4}(d) shows the SOC dependence of $\sigma_{zx}'$ at $(U, n) = (1 \ {\rm eV}, 6.2)$, where the original values of ${\bm \lambda}_{ij}$ are multiplied by a parameter $\alpha$.
In addition to the HF solutions, we plot the results without the FM moment from canting that are obtained self-consistently under the constraint of $\langle S^x \rangle=\langle S^y\rangle=0$ for comparison.
In both cases, $\sigma_{zx}'$ increases in proportion to $\alpha$, and surprisingly, the difference between them is small, only a few percent. 
This result indicates that the canted FM moment is irrelevant to the AHE, while the SOC is indispensable.

Then, we examine which component of the SOC is crucial.
Figure~\ref{fig4}(e) shows the behavior of $\sigma_{zx}'$, when the $z$ or $xy$ components of all $\bm{\lambda}_{ij}$ are set to zero by hand and the others are scaled by $\alpha$. 
When $\lambda_{ij}^x=\lambda_{ij}^y=0$, the ground state reduces to the collinear AFM structure without the canted FM moment and $\sigma_{zx}'$ shows almost the same linear dependence as in Fig.~\ref{fig4}(b).
In contrast, when $\lambda_{ij}^z=0$, $\sigma_{zx}'$ becomes constantly zero for any magnitude of $\alpha$, although the ground state is the canted AFM structure. 
The result indicates that the $z$ component of ${\bm \lambda}_{ij}$ is crucial for the AHE, while the $xy$ components are irrelevant.
\begin{figure*}[t]
\begin{center}
\includegraphics[width=1.75\columnwidth, clip]{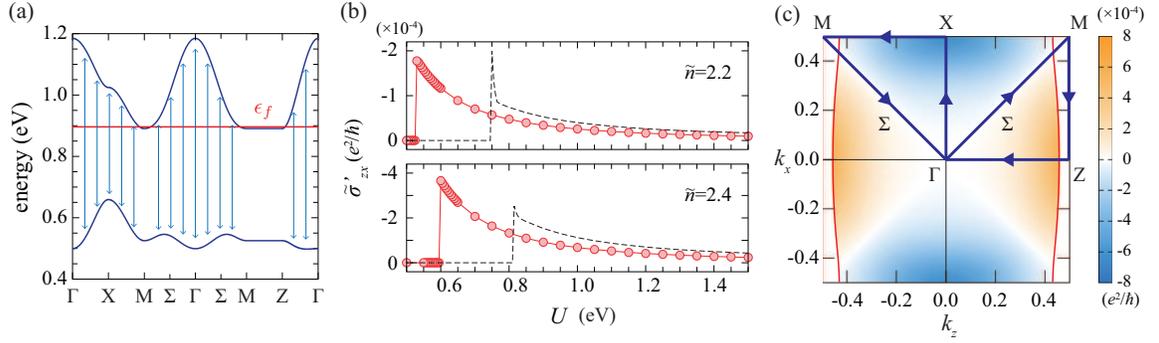}
\end{center}
\caption{
(a) Energy band structure of the effective model in the AFM metallic state at $(U, \tilde{n}) = (1 \ {\rm eV}, 2.2)$. 
$\epsilon_f$ represents the Fermi energy.
The arrows connecting the upper and lower bands shows schematically the interband transition processes contributing to the Hall conductivity. 
(b) $U$ dependences of $\tilde{\sigma}'_{zx}$ by the effective model for $\tilde{n}=2.2$ (upper panel) and $2.4$ (lower panel). 
The broken lines represent $\sigma'_{zx}$ in the original model for the same $U$ and $n=\tilde{n}+4$.
(c) Color map of the ${\bm k}$-resolved Hall conductivity $\sigma_{zx}'(\bm k)$ in the first BZ at $(U, \tilde{n}) = (1 \ {\rm eV}, 2.2)$.
The red curves show the Fermi surface. 
}
\label{fig6}
\end{figure*}

\subsection{Effective model in the strong dimerization limit}
In the following, we investigate the microscopic origin of the AHE found above, by constructing an effective model.
For this purpose, we examine how the Hall conductivity changes with the intradimer transfer integral $t_a$.
As mentioned in Sec. I, when $t_a$ is large enough, the bonding and antibonding orbitals of the dimer are energetically well separated and the fully-occupied bonding band can be neglected. 
Accordingly, the system is regarded as an effective half-filled system.
Figure~\ref{fig5}(a) shows the variation of $\sigma_{zx}'$ with respect to $t_a$ at $(U, n)=(1 \ {\rm eV}, 6.2)$. 
By increasing $t_a$, the Hall conductivity decreases but converges to a nonzero value in the limit of $t_a \rightarrow \infty$.
This implies that the essence of the AHE can be captured even in the strong dimerization limit. 

In this limit, one can construct an effective single-band model based on the antibonding orbitals of the dimers.
The transfer integrals for these orbitals are reduced to those on the anisotropic triangular lattice as shown in Fig.~\ref{fig5}(b).
As for the SOC, considering the numerical results above, we only take into account the $z$ component of the SOC vectors, i.e., the Zeeman-type SOC is considered. 
In this case, the collinear AFM state without canting is stabilized, where the AFM moments are parallel to the $z$ axis as shown in Fig.~\ref{fig5}(c).  
We note that despite these simplifications the difference between the $A$ and $B$ dimers remains in the effective model included by the stripe-like pattern of the SOC vectors as shown in Fig.~\ref{fig5}(b).

Now the effective Hamiltonian is given by
\begin{align}
{\cal H}_{\rm eff} 
&= \sum_{ij} \sum_{\sigma} \tilde{t}_{ij} \tilde{c}^{\dagger}_{i \sigma} \tilde{c}_{j \sigma} + \sum_{i \sigma} \Delta_{i} \sigma \tilde{n}_{i\sigma} \notag \\
&+ \sum_{ij} \sum_{\sigma \sigma'} \frac{i}{2} (\tilde{\lambda}_{ij} \sigma^z)_{\sigma \sigma'} \tilde{c}^{\dagger}_{i \sigma} \tilde{c}_{j \sigma'}, 
\label{heff}
\end{align}
where $\tilde{c}_{i \sigma}$ ($\tilde{c}_{i \sigma}^\dagger$) and $\tilde{n}_{i\sigma}$($=\tilde{c}_{i \sigma}^\dagger \tilde{c}_{i \sigma}$) are the annihilation (creation) and number operators of an electron on the antibonding orbital of $i$th dimer, $\tilde{t}_{ij}$ is the electron transfer integral between $i$ and $j$th dimers, $\Delta_{i}$ is the molecular field describing the collinear AFM order parallel to the $z$ axis shown in Fig.~\ref{fig5}(c), and $\tilde{\lambda}_{ij}$ is the $z$ component of the effective SOC vector. 
The coefficient of the molecular field, $\sigma$, in the second term takes $+1$ and $-1$ for up and down spins, respectively.
These parameters are given by those in the original HF Hamiltonian ${\cal H}_{\rm HF}$ as 
\begin{gather}
\tilde{t}=-\frac{1}{2}(t_p-t_q), \ \tilde{t}'=-\frac{t_b}{2}, \\
\tilde{\lambda}_{ij} = \pm \tilde{\lambda} = \pm \frac{1}{2}(\lambda_{q {\rm 1}}^z - \lambda_{p {\rm 1}}^z), 
\end{gather}
[see Fig.~\ref{fig5}(b)] and
\begin{gather}
\Delta_i = \mp \Delta = \mp \frac{\delta U}{4} 
\end{gather}
for the $A$ ($-$) and $B$ ($+$) dimer sites where $\delta = \langle \tilde{n}_{i  \in A \uparrow} \rangle - \langle \tilde{n}_{i \in A \downarrow} \rangle = \langle \tilde{n}_{i  \in B \downarrow} \rangle - \langle \tilde{n}_{i \in B \uparrow} \rangle$ determined self-consistently for each value of $U$.~\cite{note2} 

The band structure of the effective model at $(U, \tilde{n})=(1 \ {\rm eV}, 2.2)$ is shown in Fig.~\ref{fig6}(a), where $\tilde{n}$ (=$n-4$) is the number of electrons per unit cell containing two dimer sites. 
These correspond to the upper four antibonding bands in the original eight bands in Fig.~\ref{fig3}(a), separated by the energy gap due to the strong dimerization from the lower four bonding bands.
These four bands are separated into spin-degenerate two bands each by the AFM gap. 
Note the fact that in our model the unit cell has two sites owing to the SOC is in contrast with the single-band Hubbard model investigated in the previous studies.~\cite{kino, kino2, morita, tremblay, koretsune2, watanabe} 
Another point to notice is that the spin splitting is now absent and the bands are doubly degenerate in the whole BZ. 
We can understand this from the following two viewpoints.
One is that, in the strong dimerization limit, the information is lost that the $A$ and $B$ dimers are connected by the glide symmetry leading to the spin splitting by the AFM ordering, as discussed in our previous work.~\cite{naka} 
The other is because of the Zeeman-type SOC between the $A$ and $B$ dimers only keeping the $z$ component, making the model diagonal in the spin space.

As a consequence, the effective Hamiltonian is spin diagonal and block diagonalized into two $2 \times 2$ matrices.
Then we can easily diagonalize ${\cal H}_{\rm eff}$ resulting in the energies of the upper and lower bands shown in Fig.~\ref{fig6}(a) given as
\begin{align}
\epsilon^{\rm u,l}_{\bm k} = A_{\bm k} \pm \sqrt{B_{\bm k}^2 + C_{\bm k}^2 + \Delta^2},
\end{align}
where $+$ and $-$ correspond to the upper and lower branches, $\epsilon_{\bm{k}}^{\rm u}$ and $\epsilon_{\bm{k}}^{\rm l}$, respectively.
The functions $A_{\bm k}$, $B_{\bm k}$, and $C_{\bm k}$ are given by 
\begin{align}
A_{\bm k} &= 2 \tilde{t}' \cos(2 \pi k_z), \\
B_{\bm k} &= 2 \tilde{t} [\cos(\pi (k_z-k_x)) + \cos(\pi (k_z+k_x))], \\
C_{\bm k} &= - \tilde{\lambda} [\cos(\pi (k_z-k_x)) - \cos(\pi (k_z+k_x))],
\end{align}
respectively.

\subsection{Origin of the AHE}
From our effective model above, we obtain the analytic form of the real part of the Hall conductivity as 
\begin{align}
\tilde{\sigma}_{zx}' = - \frac{2e^2 \tilde{t} \tilde{\lambda} \Delta}{N \hbar} & \sum_{\bm k} \left[ f(\epsilon^{\rm u}_{\bm k}) - f(\epsilon^{\rm l}_{\bm k}) \right] \notag \\
&\times \frac{\sin [\pi (k_z - k_x)] \sin [\pi (k_z + k_x)]}{(B_{\bm k}^2 + C_{\bm k}^2 + \Delta^2)^\frac{3}{2}}.
\label{sigma_tilde}
\end{align}
Note that the damping constant is omitted in Eq.~(\ref{sigma_tilde}) since we are interested in the intrinsic contribution coming from the interband transitions as discussed above. 
Figure~\ref{fig6}(b) shows the variations of $\tilde{\sigma}_{zx}'$ as a function of $U$ at $\tilde{n}=2.2$ and $2.4$ compared to $\sigma_{zx}'$ in the original model at $n=6.2$ and $6.4$, respectively, both of which show similar $U$ and $n$ dependences.
In addition, the value of $\tilde{\sigma}_{zx}'$ well reproduces $\sigma_{zx}'$ in the strong dimerization limit obtained by the HF calculation as shown in Fig.~\ref{fig5}(a).

The analytic expression in Eq.~(\ref{sigma_tilde}) enables us to sort out the relevant parameters in the AHE. 
By neglecting the $\bm k$ dependence in Eq.~(\ref{sigma_tilde}), we find 
\begin{gather}
\tilde{\sigma}_{zx}' \propto \frac{\tilde{t} \tilde{\lambda} \Delta}{(4\tilde{t}^2+\tilde{\lambda}^2+\Delta^2)^\frac{3}{2}}.
\label{sigma_tilde2}
\end{gather}
For $\kappa$-type ET compounds, $|\Delta| > |\tilde{t}| \gg |\tilde{\lambda}|$; these parameters are typically $|\Delta| \sim 0.2$~eV, $|\tilde{t}| \sim 0.1$~eV, and $|\tilde{\lambda}| \sim 0.1$~meV.
Applying this relation to Eq.~(\ref{sigma_tilde2}), the Hall conductivity is expected to increase proportional to $\tilde{\lambda}$ and decrease with increasing $\Delta$. 
This explains the $U$, $n$, and $\lambda$ dependences of $\sigma_{zx}'$ in the original model, as was shown in Figs.~\ref{fig4}(a), \ref{fig4}(d), and \ref{fig4}(e). 
These results indicate that the essence of the present AHE is well captured by the effective model. 
Our analytic formula shows that it is attributed to the electron transfer integral $\tilde{t}$, the $z$ component of the SOC, $\tilde{\lambda}$, and the collinear AFM ordering, whose energy scale is determined by the order parameter $\delta$ multiplied by the interaction $U$. 

In addition, similar to the case for the original model, we can obtain insight into the origin of the AHE by considering its $\bm k$-resolved form as $\tilde{\sigma}_{zx}' \equiv e^2/(N \hbar) \sum_{\bm k} \tilde{\sigma}_{zx}'({\bm k})$, representing the interband electron transition processes between the lower occupied and the upper unoccupied Bloch states illustrated in Fig.~\ref{fig6}(a). 
As shown in Fig.~\ref{fig6}(c), the $d$-wave symmetry in $\tilde{\sigma}'_{zx}({\bm k})$, characterized by $\sin [\pi (k_z - k_x)] \sin [\pi (k_z + k_x)]$ in Eq.~(\ref{sigma_tilde}), is clearly seen. 
Then, as discussed for the original model, in the AFM metallic phase the interband processes on the $\bm k$ points outside the Fermi surface are excluded from the summation in Eq.~(\ref{sigma_tilde}) by the Fermi distribution function $f(\epsilon^{\rm u}_{\bm k}) - f(\epsilon^{\rm l}_{\bm k})$, which leads to the nonzero Hall conductivity in the AFM metallic phase.

\begin{figure}[t]
\begin{center}
\includegraphics[width=1.0\columnwidth, clip]{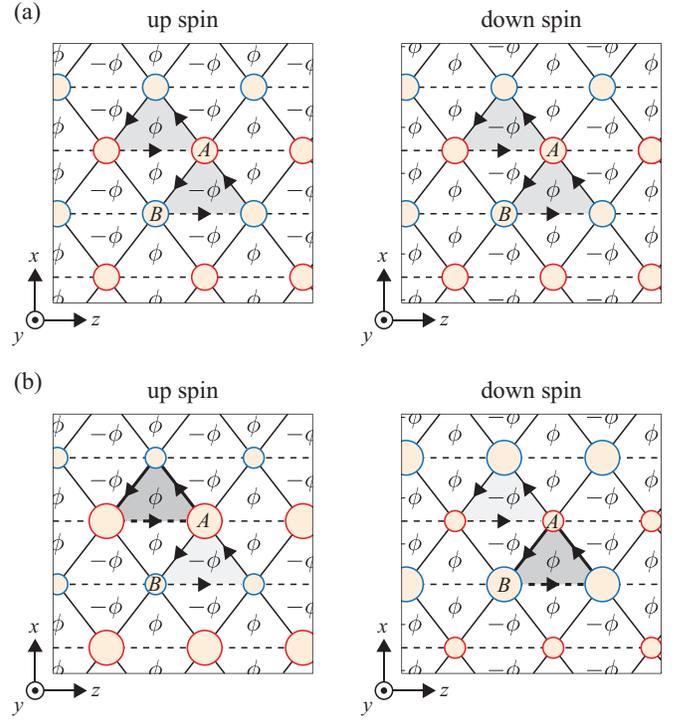}
\end{center}
\caption{
Real-space distribution of the fictitious magnetic fluxes acting on the conduction electrons with up spin (left panel) and down spin (right panel) in the PM metallic phase (a) and the AFM metallic phase (b). 
The solid and broken arrows represent the directions of the line integral defining the magnetic flux inside the loop for the electron hoppings $\tilde{t}$ and $\tilde{t}'$, respectively.
The shaded triangles represent the upward ($ABA$ and $BAB$) triangles with counterclockwise direction of the loop integrals. 
They are equivalent in the PM phase (a) but inequivalent by the spin-dependent charge imbalance in the AFM phase (b), as shown by the different shading.  
The large and small circles in the left (right) panel in (b) represent the up (down)-spin-rich and poor dimer sites in the AFM phase, respectively.
}
\label{fig7}
\end{figure}
\subsection{Real-space picture}
Here, we provide an intuitive picture of the AHE in terms of the real-space fictitious magnetic fields acting on the conduction electrons.~\cite{ohgushi, zhang}
They are given through the SOC term in the effective model on the anisotropic triangular lattice composed of the dimer sites. 
We define the magnetic flux $\psi$ penetrating each basic triangle in the lattice as $\exp(i \psi) = \exp(i \oint_{C} {\bm A} \cdot d{\bm c})$, where $\bm A$ represents the vector potential associated with the path $C$ along the three sides of the triangles in the counterclockwise direction as shown in Fig.~\ref{fig7} (only the paths for the upward triangles are shown). 
The complex transfer integrals between the $A$ and $B$ dimers on these paths are written as $\tilde{t} \mp i \tilde{\lambda} \sigma/2 = r \exp (\pm i \theta \sigma)$, where $\sigma$ represents the $z$ component of the spin and the upper and lower signs corresponds to the $ABA$ and $BAB$ triangles, respectively. 
Note that paths along $\tilde{t}'$ do not contribute due to the absence of SOC. 
The magnetic fluxes acting on the up-spin electron rotating the upward $ABA$ and $BAB$ triangles are given by $\psi = \pm \phi = \pm 2\theta$, while those for the down spin electron are given by $\psi = \mp \phi = \mp 2\theta$. 
Thus, the real-space distributions of the fluxes for the up- and down-spin electrons are given by the staggered arrangements with opposite signs both in the PM and AFM metallic states as shown in Fig.~\ref{fig7}. 

In the PM state, as the up and down spin electrons are uniformly distributed on the $A$ and $B$ dimers as shown in Fig.~\ref{fig7}(a), the $ABA$ and $BAB$ loops are equivalent and the conduction electrons experience the $+\phi$ and $-\phi$ fluxes equally. 
This results in the cancellation of the net magnetic field and the zero Hall conductivity in the PM phase.

On the other hand, in the AFM state, although the flux distributions are the same as in the PM phase, the electron densities are spin dependent and form stripe-like patterns as shown in Fig.~\ref{fig7}(b); the up- and down-spin electrons accumulate more on the $A$ and $B$ dimers, respectively. 
Owing to this imbalance, the cancellation of the magnetic fluxes becomes incomplete and both the up- and down-spin electrons 
feel more the $+\phi$ fluxes, because in both up and down spin cases, all the $+\phi$ fluxes are surrounded by the two electron-rich dimers while the $-\phi$ fluxes are by the two electron-poor dimers as shown in Fig.~\ref{fig7}(b).
Consequently, both up- and down-spin conduction electrons driven by the electric field experience a net magnetic field and drift to the same direction, which results in a nonzero Hall conductivity.
These considerations lead us to an intuitive understanding of the origin of the AHE: the interplay of the staggered magnetic fluxes due to the SOC and the staggered spin-dependent electron densities owing to the AFM ordering.

\subsection{Optical AHE}
\begin{figure}[t]
\begin{center}
\includegraphics[width=1.0\columnwidth, clip]{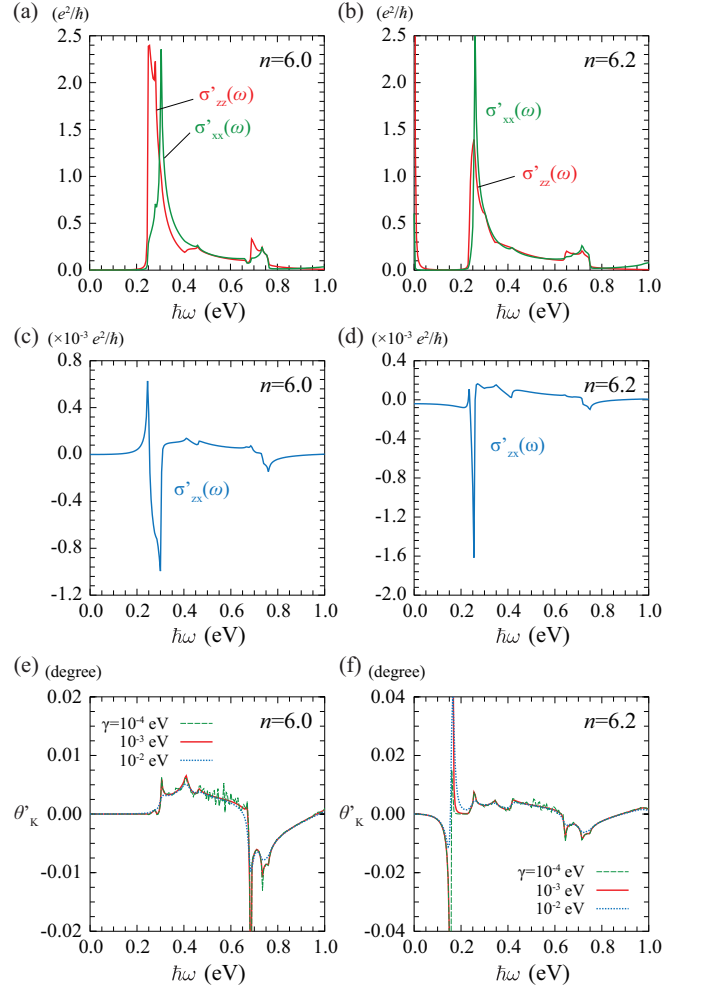}
\end{center}
\caption{
(a) Optical conductivity spectra $\sigma_{zz}'(\omega)$ and $\sigma_{xx}'(\omega)$, (c) optical Hall conductivity spectrum $\sigma_{zx}'(\omega)$, and (e) magneto-optical Kerr rotation angle $\theta_{\rm K}'(\omega)$ in the AFM insulating phase at $(U, n)=(1 \ {\rm eV}, 6)$.
Corresponding data for the AFM metallic phase at $(U, n)=(1 \ {\rm eV}, 6.2)$ are shown in (b), (d), and (f).
The damping constant is fix at $\gamma=1$ meV in (a), (b), (c), and (d).
}
\label{fig8}
\end{figure}

Finally, we examine the optical responses with nonzero frequency $\omega$ under the AFM ordering. 
First, we show the results of the longitudinal optical conductivity spectra calculated for the original HF Hamiltonian ${\cal H}_{\rm HF}$ in the AFM insulating phase at $(U, n) = (1.0 \ {\rm eV}, 6)$ in Fig.~\ref{fig8}(a). 
In the longitudinal conductivities, there are two peak structures around $\hbar \omega=0.3$ eV and $0.7$ eV.
The origins of these peaks have been discussed in previous studies;~\cite{kezmarki, naka3, naka2, hashimoto} the lower and higher energy peaks are identified as the charge transfer excitations between the antibonding orbitals of the neighboring dimers and the intradimer charge excitation from the bonding to antibonding orbital, respectively.
In the doped AFM metallic phase at $(U, n)=(1 \ {\rm eV}, 6.2)$, the Drude weight appears in the optical conductivity spectra at $\hbar \omega=0$ as shown in Fig.~\ref{fig8}(b).

Next, we show the results of the optical Hall conductivity in the AFM insulating phase at $(U, n) = (1 \ {\rm eV}, 6)$ in Fig.~\ref{fig8}(c). 
The optical Hall conductivity shows nonzero oscillator strength between the two peak energies, while it approaches zero toward $\hbar\omega=0$.
On the other hand, in the AFM metallic phase at $(U, n)=(1 \ {\rm eV}, 6.2)$, the Hall conductivity shows similar behavior, while it has nonzero DC component as shown in Fig.~\ref{fig8}(d).
These results suggest that for a linearly polarized light incident perpendicularly on the $zx$ plane, the reflected light is ellipsoidally polarized, i.e., the magneto-optical Kerr effect occurs. 

The magneto-optical Kerr rotation angle is given in the complex form
\begin{align}
\theta_{\rm K}(\omega) 
&= \theta_{\rm K}'(\omega) + i \theta_{\rm K}''(\omega) \\ 
&= \frac{-\sigma^{\rm 3D}_{zx}(\omega)}{\sigma^{\rm 3D}_{zz}(\omega) \sqrt{1+i \sigma^{\rm 3D}_{zz}(\omega) / (\omega \epsilon_0)}},
\end{align}
when the incident light is polarized parallel to the $z$ axis and the angle is small.~\cite{argyres, reim}
$\epsilon_0$ is the permittivity of vacuum.
$\sigma_{\mu\nu}^{\rm 3D}(\omega)$ is the bulk electrical conductivity tensor for the three-dimensional system, which is related to the electrical conductivity tensor in the two-dimensional system as $\sigma_{\mu\nu}^{\rm 3D}(\omega) = \sigma_{\mu\nu}(\omega)/d$, where $d$ is the distance between the neighboring layers and typically given by $\sim 15$ ${\rm \AA}$ for the $\kappa$-ET systems.~\cite{williams}
The $\omega$ dependences of the real part, $\theta_{\rm K}'$, in the AFM insulating and metallic phases are presented in Figs.~\ref{fig8}(e) and \ref{fig8}(f), respectively. 
The magnitudes of the sharp peak structures change depending on the damping constant $\gamma$, while the overall behaviors are insensitive to $\gamma$.
The absolute values of the rotation angles are enhanced between the interdimer and intradimer charge excitation energies and reach about $0.04$ degree.

\section{Discussion}
Let us discuss experimental measurements of the present AHE in the $\kappa$-type ET compounds, especially, the two representative compounds showing AFM ordering, $\kappa$-(ET)$_2$Cu[N(CN)$_2$]Cl (abbreviated as h-Cl) and the deutrated $\kappa$-(ET)$_2$Cu[N(CN)$_2$]Br (abbreviated as d-Br); ``h'' and ``d'' represent hydrogen and deuterium atoms in the ethylene groups on both ends of an ET molecule, respectively.
According to recent experiments, in both compounds, the AFM spin structure in the ET layer is common as shown in Fig.~\ref{fig1}(b), while the interlayer stacking structure is different with each other.~\cite{ishikawa, taniguchi} 
In h-Cl, the neighboring $A$ ($B$) dimers along the $y$ axis has the same sign of $\langle S^z \rangle$, i.e., an ``in-phase'' stacking of the AFM ordering is realized, while it is ``anti-phase'' in d-Br. 
This difference is crucially important for the observation of the AHE in the bulk compounds, as follows. 

Based on the analytic form of the Hall conductivity in Eq.~(\ref{sigma_tilde}), the sign of the AHE in a single ET layer is determined by the sign of the product $\tilde{t} \Delta \tilde{\lambda}$.
In the actual three-dimensionally stacked ET systems, the AFM order parameter $\Delta$ can take different signs between the layers, depending on the stacking manner of the AFM ordered structure. 
The above AFM structures realized in h-Cl and d-Br are interpreted as the states in which $\Delta$ of each ET layer is arranged uniformly and alternately along the interlayer $y$ direction, respectively.
On the other hand, the SOC $\tilde{\lambda}$ reflects the symmetry of the interlayer molecular arrangement. 
h-Cl and d-Br belong to the same space group {\it Pnma} with a mirror plane between the neighboring ET layers perpendicular to the $y$ axis. 
By this mirror symmetry, the sign of $\tilde{\lambda}$ reverses between the neighboring layers in both compounds.
Therefore, considering the relative signs of $\tilde{t} \Delta \tilde{\lambda}$ for the ET layers, the net AHE is expected to survive only in d-Br while it will be cancelled out in h-Cl. 
This prediction provides a good testbed for our scenario in experiments.
In addition, the energy difference between the two kinds of the AFM structures is quite small, and therefore they can be easily inverted by applying a magnetic field. 
In fact, the AFM structure of d-Br has been obtained also in h-Cl by applying the magnetic field of $5$ Tesla along the $y$ axis.~\cite{ishikawa} 
Taking advantage of this property, one might be able to not only examine the AHE without comparing the two different compounds but also toggle on and off of the AHE by the magnetic field in h-Cl.

The observation of the DC AHE shown in Fig.~\ref{fig4}(a) basically requires carrier doping to the $\kappa$-type ET systems. 
Recently, carrier-doping techniques to the organic compounds have rapidly been developed using anion substitutions and electrical double layer devices where a doping-induced Mott transition and superconductivity have actually been reported.~\cite{kawasugi, kawasugi2, kawasugi3, taniguchi2, oike} 
Based on these advances in experiments, a verification of the AHE in the doped Mott insulators is expected to be feasible in the near future. 
In turn, such experiments can provide important information about the magnetic state in the doped organic Mott insulators, which is sometimes difficult to identify in thin film samples by technical reasons. 
Besides, the observation of the magneto-optical Kerr effect is another promising way to investigate the present mechanism. 

Most recently, a possibility of the AHE in collinear antiferromegnets has been explored theoretically and experimentally in inorganic compounds with the rutile structure, e.g., RuO$_2$ and NiF$_2$.~\cite{smejkal, li, feng}
However, studies for the AHE in inorganic compounds often involve experimental difficulties in isolating the intrinsic contribution purely attributed to the electronic band structure, from the extrinsic contributions due to impurities. 
Furthermore, their complicated band structures sometimes prevent from extracting the key ingredients theoretically. 
In contrast, organic crystals generally contain less impurities and have a simpler band structure due to the low-symmetric molecules leading to the energetically-isolated frontier orbitals than inorganics, which provide an ideal platform on studying the intrinsic AHE.

\section{Summary}
We have proposed the possibility of the AHE in organic antiferromagnets with the $\kappa$-type molecular arrangement. 
The present AHE originates from not the FM moment by spin canting but the collinear AFM ordering, in contrast to the conventional AHE in ferromagnets and noncollinear antiferromagnets.
The microscopic origin is the cooperation of the staggered fictitious magnetic field emerging from the SOC incorporated in the molecular arrangement and the spin-dependent electron density owing to the collinear AFM ordering, which results in the net Lorenz force acting on the conduction electrons.
Our scenario can be verified by comparing the DC AHE and the optical AHE in the $\kappa$-type ET compounds showing the different types of the AFM ordering structures, $\kappa$-(ET)$_2$Cu[N(CN)$_2$]Cl and the deutrated $\kappa$-(ET)$_2$Cu[N(CN)$_2$]Br. 

MN and HS would like to thank H. Fukuyama, T. Furukawa, S. Iguchi, K. Riedl, T. Sasaki, H. Taniguchi, and S. M. Winter for valuable comments and discussions. 
This work is supported by Grant-in-Aid for Scientific Research, No. JP16K17731, No. JP19K03723, No. JP18H04296 (J-Physics), No. JP18K13488, No. JP15H05885(J-Physics), No. JP19K03752, No. JP19K21860, JP19H05825, JST-CREST(No. JPMJCR18T2), and the GIMRT Program of the Institute for Materials Research, Tohoku University, No. 19K0019.



\begin{thebibliography}{99}
\bibitem{note4}
ET is the abbreviation of BEDT-TTF, bis(ethylenedithio)tetrathia-fulvalene molecule.
\bibitem{note3}
In our previous paper we took the $xy$ plane in the conducting ET layer.
\bibitem{note}
When $n$ decreases, i.e., in the case of hole doping, a ferrimagnetic metallic state with a small charge disproportionation between the $A$ and $B$ dimers is stabilized in a wide range of parameters.
\bibitem{note2}
The mean fields in the strong dimerization limit are calculated by the original model with $t_a \rightarrow \beta t_a$ and $\beta = 10^3$, in which these values are converged enough with respect to $\beta$.
\end{thebibliography}
\end{document}